\documentclass[fleqn,usenatbib]{mnras}

\usepackage{newtxtext,newtxmath}
\usepackage[T1]{fontenc}
\usepackage{ae,aecompl}

\usepackage{graphicx}    % Including figure files
\usepackage{amsmath}    % Advanced maths commands
\usepackage{amssymb}    % Extra maths symbols

\usepackage{times}

\newcommand{\MSun}{\mbox{${\rm M}_\odot$}}

\def\lteq{\ {\raise-.5ex\hbox{$\buildrel<\over-$}}\ }
\def\apgt{\ {\raise-.5ex\hbox{$\buildrel>\over\sim$}}\ }
\def\aplt{\ {\raise-.5ex\hbox{$\buildrel<\over\sim$}}\ }
\def\lt{\ {\raise-.5ex\hbox{$\buildrel>$}}\ }
\def\gt{\ {\raise-.5ex\hbox{$\buildrel<$}}\ }
\def\eqgt{\ {\raise-.5ex\hbox{$\buildrel>\over-$}}\ }
\def\eqlt{\ {\raise-.5ex\hbox{$\buildrel<\over-$}}\ }

\newcommand{\A}{'Oumuamua\,}
\newcommand{\solus}{\emph{s\={o}lus lapis}}
\newcommand{\soli}{\emph{s\={o}l\={\i} lapid\={e}s}}
\title[Rendezvous with 'Oumuamua]
      {The origin of interstellar asteroidal objects like 1I/2017 U1 'Oumuamua}

\author[S. Portegies Zwart et al.]
{S.\, Portegies Zwart$^{1}$\thanks{Emain: spz@strw.leidenuniv.nl},
 S.\, Torres$^{1}$, 
 I.\, Pelupessy$^{2}$,
 J. B\'edorf$^{1}$,
 Maxwell\, X.\, Cai$^{1}$ \\
$^{1}$ Leiden Observatory, Leiden University, NL-2300RA Leiden, The Netherlands\\
$^{2}$ The Netherlands eScience Center, The Netherlands
 }

\date{Accepted XXX. Received YYY; in original form ZZZ}

\pubyear{2015}

\begin{document}
\label{firstpage}
\pagerange{\pageref{firstpage}--\pageref{lastpage}}
\maketitle

\begin{abstract}
  
We study the origin of the interstellar object 1I/2017 U1 \A by
juxtaposing estimates based on the observations with simulations.  We
speculate that objects like \A are formed in the debris disc as left
over from the star and planet formation process, and subsequently
liberated. The liberation process is mediated either by interaction
with other stars in the parental star-cluster, by resonant
interactions within the planetesimal disc or by the relatively sudden
mass loss when the host star becomes a compact object.  Integrating \A
backward in time in the Galactic potential together with stars from
the Gaia-TGAS catalogue we find that about 1.3\,Myr ago \A passed the
nearby star HIP 17288 within a mean distance of $1.3$\,pc.  By
comparing nearby observed L-dwarfs with simulations of the Galaxy we
conclude that the kinematics of \A is consistent with relatively young
objects of $1.1$--$1.7$\,Gyr.  We just met \A by chance, and with a
derived mean Galactic density of $\sim 3\times 10^{5}$ similarly sized
objects within 100\,au from the Sun or $\sim 10^{14}$ per cubic parsec
we expect about 2 to 12 such visitors per year within 1\,au from the
Sun.

\end{abstract}

\begin{keywords}
  methods: data analysis ---
  methods: statistical ---
  methods: numerical  ---
  methods: observational ---
  minor planets, asteroids: individual: 'Oumuamua  ---
  Galaxy: local interstellar matter
\end{keywords}

\section{Introduction}

1I/2017~U1~'Oumuamua (hereafter \A), a genuine interstellar object
that was discovered on 19 October 2017 in the Pan-STARRS survey
\citep{2017MPEC....U..181B,2017MPEC....U..183M,2017MPEC....W...52M},
%% \citep{A2017U1_U181,A2017U1_U183},
was initially identified as a comet but quickly reclassified as an
{\em unusual minor planet}.  Most notable orbital parameters are the
distance at pericentre, $q = 0.25534 \pm 0.00007$\,au, the eccentricity,
$e = 1.1995 \pm 0.0002$\, and the relative velocity at infinity $v
\simeq 26.32 \pm 0.01$\,km/s with respect to the Sun
\citep{2017MPEC....U..181B,2017MPEC....U..183M,2017RNAAS...1...21M}. This
makes the object unambiguously unbound on an en-passant orbit through
the Solar system \citep{2017RNAAS...1....5D,2017ApJ...851L..38B}, not
unlike the prediction by \cite*{2009ApJ...704..733M}. The absence of a
tail indicates that the object is probably rock-like
\citep{2018NatAs...2..112M}, which, together with its unusual
elongated shape \citep[$\sim 35$ by 230\,m reported by][and explained
  by \cite{2018NatAs...2..133F} and
  \cite{2017RNAAS...1...50D}]{2017Natur.552..378M} and rapid spin
\citep[with an $8.10\pm0.02$ hour period,
  see][]{2017ApJ...850L..36J,2018ApJ...852L...2B}, has been hosted by
a star for an extensive period of time
\citep{2018arXiv180202273K,2018arXiv180201335H}. In this period it may
have lost most of its volatiles and ice by sputtering
\citep{2018NatAs...2..112M,2018NatAs...2..133F}.  By the lack of other
terminology for this class of objects and for convenience, we refer in
this work to \solus{}, which is Latin for ``lonely stone''.

Spectra taken by several instruments indicate that its colour ($g-r =
0.2 \pm 0.4$ and $r-i = 0.3\pm 0.3$ \citep{2017arXiv171009977M}, $g-r
=0.41\pm 0.24$ and $r-i=0.23 \pm0.25$ \citep{2018ApJ...852L...2B}, and
$g'-r' = 0.60 \pm0.23$ \citep{2017ApJ...851L...5Y}) matches that of
objects in the Kuiper belt \citep{1951PNAS...37....1K} but that it is
somewhat red compared to the main belt asteroids and trojans \cite[see
  also][]{2017ApJ...851L..38B,2018AJ....155...56J}.  However, it
cannot be excluded that \A is the left over nucleus of a comet from
another star
\citep{2017arXiv171107535F,2018NatAs...2..112M,2018MNRAS.tmp..493R}.

Based on its velocity, \A is unlikely to have originated
from a nearby star
\citep{2017arXiv171206721G,2017RNAAS...1...21M,2017RNAAS...1....5D},
although \cite*{2017RNAAS...1...13G} argue that it could have been
ejected with a low velocity (of $1$--$2$\,km/s) from the nearby young
stellar associations Carina or Columba or from the Pleiades moving
group \citep{2018ApJ...852L..27F}.  Tracing back the trajectory of
\A\, through the local neighbourhood is hindered by the magnification
of small uncertainties in its orbit and in the positions and
velocities of nearby stars \citep{2018ApJ...852L..13Z}.  Even with the
high accuracy achieved by Gaia and TGAS, backtracing its orbit can
only be done reliably for some 10\,Myr or over a distance of about
30\,pc \citep{2017RNAAS...1...21M}.

Since we doubt that \A formed as an isolated object, it seems most
plausible that it formed around another star, from which it was
liberated.  This could be the result of copious mass loss resulting
from the end phase of the nuclear burning of a star
\citep{2011MNRAS.417.2104V,2017RNAAS...1...55H,2018arXiv180102658R},
by a close encounter in a young star cluster
\citep{2016MNRAS.457.4218J}, or by internal scattering of planetary
systems \citep*{2013Icar..225...40B,2018arXiv180302840R} or in a binary
star \citep{2017arXiv171204435J,2018ApJ...852L..15C}.  It is even
possible that one of the Sun's own Oort cloud objects was scattered by
a yet unknown planet \citep*{2017RNAAS...1...38W,2018MNRAS.476L...1D}.

Here we discuss the possible origin of \A (see
\S\,\ref{Sect:Aisrare}). We analyse its orbital properties and compare
those with our expectations for a random population of free-floating
debris in the solar neighbourhood (see \S\,\ref{Sect:Origin}).  Our
estimates of the occurrence of objects similar to \A\, is juxtaposed
with several alternative scenarios.  In the near future when the Large
Synoptic Telescope comes online many more \soli{} are expected to be
found at a rate of 2 to 12 per year.

\section{'Oumuamua as a rare object}\label{Sect:Aisrare}

\subsection{Observational constraints}\label{Sect:ObsRate}

A simple estimate of the number density of \soli{} implied by the
detection of \A can be derived by calculating the effective volume
surveyed by the Pan-STARRS telescope \citep[A much more elaborate
  analysis was recently presented in][]{2018ApJ...855L..10D}. This
telescope has a limiting magnitude of $m \sim 22$
\citep{2017MPEC....U..181B,2017MPEC....U..183M}. \A was observed with
a magnitude $H = 20.19$ and was within a distance of about $0.16$\,au
from Earth before being detected.\footnote{The quoted value for $H$ is
  the absolute magnitude which is defined as the magnitude at 1\,au
  from both the Earth and the Sun.}

The volume surveyed by the Pan-STARRS telescope is approximately:
\begin{equation}
  V \simeq A v \Delta t_{\rm PAN}.
  \label{Eq:Volume}
\end{equation}
Here $A$ is the effective cross section for the object passing the Solar system:
\begin{equation}
  A = \pi r_{\rm detect}^2 \left(1 + \left(v_{\rm esc} \over v \right)^2 \right).
  \label{Eq:Crosssection}
\end{equation}
In which $\pi r_{\rm detect}^2$ is the geometric cross section and the
second term corrects for the gravitational focusing.  Here $v$ is the
relative velocity of \A at infinity with respect to the Solar system's
barycentre and $v_{\rm esc} \simeq 42$\,km/s is the escape velocity of
the Sun at a distance of 1\,au.  The period over which Pan-STARRS has
observed $\Delta t_{\rm PAN} \simeq 5$\,yr \citep[the first data
  release in 2014 covers 3 years, see][]{2017AJ....153..133E}.  When
we naively assume that the telescope has a 100\,\% detection
efficiency over these 5\,years we arrive at a density of $\sim
0.08$\,au$^{-3}$ or $\sim 7.0 \times 10^{14}$\,pc$^{-3}$.  The
statistical uncertainty implied by a single detection means that a
wide range of values ($0.004$--$0.24$\,au$^{-3}$ or $3.5 \times
10^{13}$--$2.1 \times 10^{15}$\,pc$^{-3}$) is consistent with the
observations at the $95\%$ level.

This simple estimate ignores the modeling of the detection efficiency.
In spite of this, we obtain a value higher than the upper limits for
finding \soli{} derived from the absence of detections using the
Pan-STARSS1 survey up to that time by \cite{2017AJ....153..133E}.
Their estimates lead to a lower density because they considered larger
objects with a cometary tail that could have been detected to a much
larger distance from the Earth.  We will now discuss the possible
origin of \soli{} and their relation to \A.

\subsection{Dynamical ejection from the Sun's Oort cloud}\label{Sect:OortCloud}

Considering the orbital parameters of \A, it seems unlikely to have
been launched from the Kuiper belt or Oort cloud
\citep{1950BAN....11...91O}.  Asteroids and comets are frequently
launched from the Kuiper belt and the Oort cloud but they rarely
penetrate the inner Solar system \citep{2018MNRAS.476L...1D}.  Objects
in the Oort cloud, however, could possibly launch into the inner Solar
system after a strong interaction with a passing star
\citep{2008CeMDA.102..111R} or a planet \citep{2014Icar..231...99F}.

The highest velocity, in this case, is obtained when the object would
have been the member of a close binary with another, more massive,
rocky object; much in the same way in which hypervelocity stars are
ejected from the supermassive black hole in the Galactic centre
\citep{1988Natur.331..687H}. Binaries are known to be common in the
outer regions in the Solar system, and possibly the majority of
planetesimals in the Kuiper belt formed as pairs
\citep{2017NatAs...1E..88F}.  The mean orbital separation of 13
trans-Neptunian objects is $13900 \pm 14000$\,km
\citep{2008ssbn.book..345N} with a minimum of $1830$\,km for the pair
Ceto/Phorcys (object \#65489), but there is no particular reason why
Oort cloud or Kuiper-belt binaries could not have tighter orbits
\citep{2017NatAs...1E..88F}, such as is suspected of 2001~QG298
\citep{2004AJ....127.3023S}.  An interaction of an object much like
the binary (136108) 2003 EL61 \citep{2005IAUC.8636....1B} with a
planet with ten times the mass of Jupiter could result in a runaway
velocity of the least massive member of $\sim 1$\,km/s, but with the
parameters for Ceto/Phorcys the mean kick velocity of $\sim
0.13$\,km/s. This would be sufficient to change a circular orbit at a
distance of $1000$--$10^4$\,au around the Sun into an unbound orbit
with the eccentricity observed for \A.

We perform Monte Carlo experiments in which we distribute Oort cloud
objects around the Sun using the parameters adopted in
\cite{2018MNRAS.473.5432H}. We introduce a velocity kick $v_{\rm
  kick}$, taken randomly from a Gaussian distribution with dispersion
of 0.13\,km/s in each of the Cartesian directions, which results in a
wide distribution in semi-major axis and eccentricity.  More than 85\%
of the objects remain bound and the remaining 15\% unbound objects
tend to have high eccentricities with a median of $\sim 1.2$, but only
a very small fraction of $\aplt 10^{-3}$ objects pass the Sun within
100\,au on their unbound orbit \citep[see
  also][]{2017Ap&SS.362..198D}.  We subsequently performed several
direct $N$-body simulations to verify this result. For these we
adopted the connected-component symplectic integrator {\tt Huayno}
\citep*{2014A&A...570A..20J} within the {\tt AMUSE} \citep{AMUSE}
framework. The mild kicks expected due to an interaction with an Oort
cloud binary and a massive planet, either passing through or bound to
the Oort cloud, are insufficient to explain the observed velocity of
\A. We, therefore, conclude that \A is unlikely to have originated
from either the Kuiper belt or the Oort cloud \citep[see
  also][]{2018MNRAS.476L...1D}.

\subsection{Origin from the Oort cloud of another star}\label{Sect:OtherOortCloud}

Naive estimates of the density of interstellar comets can be derived
from the stellar density and assuming that each star has a rich Oort
cloud with $N_{\rm comets} \sim 10^{11}$
comets per star \citep{2009Sci...325.1234K}
\begin{equation}
  n_{\rm comets} = \eta n_{\rm stars} N_{\rm comets}.
\end{equation}
With a mean mass density in the solar neighbourhood of $\rho_0 =
0.119_{-0.012}^{+0.015}$\,\MSun/pc$^{3}$ \citep{2017arXiv171107504W}
\citep*[or $0.09\pm0.02$\,\MSun/pc$^{3}$ according
  to][]{2018MNRAS.473.2188K} and a mean stellar mass $\sim
0.37$\,\MSun\, \citep{2003PASP..115..763C}, we arrive at a local
stellar number density in the Galactic disc of $n_{\rm stars} \simeq
0.32$\,pc$^{-3}$. The efficiency at which comets are ejected from a
star when it orbits the Galaxy was recently estimated to be $\eta \sim
0.1$/Gyr \citep{2018MNRAS.473.5432H}.  With a mean stellar age of
6\,Gyr, we arrive at a local comet density of $n_{\rm comets} \sim 2.5
\times 10^{-6}$\,au$^{-3}$ (or $2.2 \times 10^{10}$\,pc$^{-3}$), which
is consistent with the upper limit on the density of interstellar
comets of $\aplt 0.0007$\,au$^{-3}$ by \cite{1990PASP..102..793S} and
$<0.001$\,au$^{-3}$ by \cite{1989ApJ...346L.105M} and
\cite{2005ApJ...635.1348F}.  Here we did not correct for the intrinsic
size and mass distribution of Oort cloud objects.  The density of
\soli{} that originate from ejected Oort-cloud objects are
considerably lower than our expected density based on the observations
derived in \S\,\ref{Sect:ObsRate}, and we conclude that \A cannot have
originated from the hypothetical exo-Oort cloud of other stars.

Copious loss of circumstellar material, however, can be initiated when
a star turns into a compact remnant either on the post-asymptotic
giant branch \citep{2011MNRAS.417.2104V} or in a supernova
\citep{1961BAN....15..291B}.  If each of the $2.0\times 10^{9}$ white
dwarfs in the Galactic disc \citep{2009JPhCS.172a2004N} has produced
${\cal O}(10^{11})$ free-floating objects, the number of \soli{} would
be over-produced by $\sim 5$ orders of magnitude.

\subsection{Origin from the debris disc of another star}

A young planetary system with a circumstellar disc may be rather rich
in relatively large $\apgt 100$\,m objects, because it is the expected
equilibrium size for collisional cascade of material-strength
dominated bodies \citep*{2013AJ....146...36S}.  Disruption of such a
disc may inject a large number of these objects into interstellar
space. This would happen in the early evolution of the star when it
still was a member of its parental cluster.  A similar estimate as in
\S\,\ref{Sect:OtherOortCloud} then reveals a total mass of ejected
disc-material to be
\begin{equation}
m_{\rm debris} \sim f_{\rm debris} Z \rho_0
\end{equation}
When we adopt a disc mass $M_{\rm disc} = 0.01 M_\odot$, and a
metallicity of the disc, $Z = 0.02$, the fraction of expelled material
is harder to estimate but it may be in the range of $f_{\rm debris}
\sim 0.1$. This fraction could even be higher if we argue that the
Solar system may have lost a fraction $\apgt 10$\% of its disc in a
resonance interaction between the outer most planets
\citep{2005Natur.435..466G}.  The total mass of ejected metals is then
of the order $\sim 2.4\times 10^{-6}$\,\MSun/pc$^{-3}$. The density of
\A\, is estimated to be $\sim 2.0$\,g/cm$^3$ \citep[see
  e.g.][]{2017ApJ...850L..38T}, which is in the range of densities
observed for the relatively ordinary Kuiper-belt object Orcus
\citep[$\rho_{\rm KBO} = 1.65^{+0.34}_{-0.24}$\,g/cm$^3$, see
  e.g.][]{2017AJ....154...19B} and the dense metal-rich 16~Psyche
\citep[$4.5 \pm 1.4$g/cm$^3$, see e.g.][]{2017Icar..281..388S}.  With
the observed dimensions of a cylinder of $25 \times 230$\,m
\citep{2017ApJ...851L..31K}, we arrive at a mass of $m \simeq 1.2
\times 10^{9}$\,kg.  A hypothetical young Solar system could then have
ejected $3.3 \times 10^{16}$ objects during its early evolution,
leading to a mean density of $\sim 3.9 \times 10^{15}$\,pc$^{-3}$,
exceeding our observational estimate by a factor of $\sim 6$.  These
objects may be cometary or non-cometary in nature depending on whether
these are ejected from inside or outside the snow line.  The large
uncertainties in this estimate, and those in
\S\,\ref{Sect:OtherOortCloud} make it plausible that an enormous
population of expelled objects exists in the Galaxy with a density
consistent with, or even exceeding our number-density estimate in
\S\,\ref{Sect:ObsRate}.

\section{Where did 'Oumuamua come from}\label{Sect:Origin}

Having compared the number of interstellar objects produced by
asteroidal or cometary ejection in \S\,\ref{Sect:Aisrare} it seems
plausible that \A was formed around another star, and was liberated
upon either close planetary encounters within the young planetary
system, or due to the copious mass loss when the star became a compact
remnant.  We now discuss the possible origin of \A.

\subsection{An origin from the solar neighbourhood}

Results by the \cite{2016A&A...595A...2G} provide the most complete
and accurate census of the distribution of stars in the solar
neighbourhood (for which we adopt a volume centred around the Sun with
a radius of 50\,pc). The radial velocity components are not (yet) part
of this database, but they can be completed using other catalogues. By
matching the TGAS catalogue with the radial velocities from
\cite{2017AJ....153...75K}, Pulkovo \citep{2006AstL...32..759G}, and
Geneva-Copenhagen \citep*{2009A&A...501..941H}, we constructed a
catalogue of 270,664 stars with position and velocity information
\citep*{2017IAUS..330T}, we selected those stars with relative proper
motions and parallax errors $<1$\,\%, with radial velocity
$<100$\,km/s, with errors $<10$\,km/s \citep[see
  also][]{2018ApJ...852L..27F}.

The orbits of the selected stars as well as \A are integrated backwards
in time for 10\,Myr in the Galactic potential.  The uncertainty in the
Gaia database and in the orbital parameters of \A are too large to
integrate reliably backwards in time any further \citep[see also][who
  adopted similar criteria]{2017RNAAS...1...21M}.  The integration was
performed using {\tt Galaxia}~\citep*{2014A&A...563A..60A}.  This
semi-analytic Milky Way Galaxy model is incorporated in the {\tt
  AMUSE} software framework
\citep*{2013CoPhC.184..456P,2013AA...557A..84P,AMUSE} using the
parameters derived by \cite*{2015MNRAS.446..823M}.  As position for the
Sun, we adopted $x=8300$\,pc, with a $z$ component of 27\,pc, and a
velocity vector of (11.1, 232.24, 7.25)\,km/s.

\begin{figure}
\begin{center}
\includegraphics[width=1.0\columnwidth]{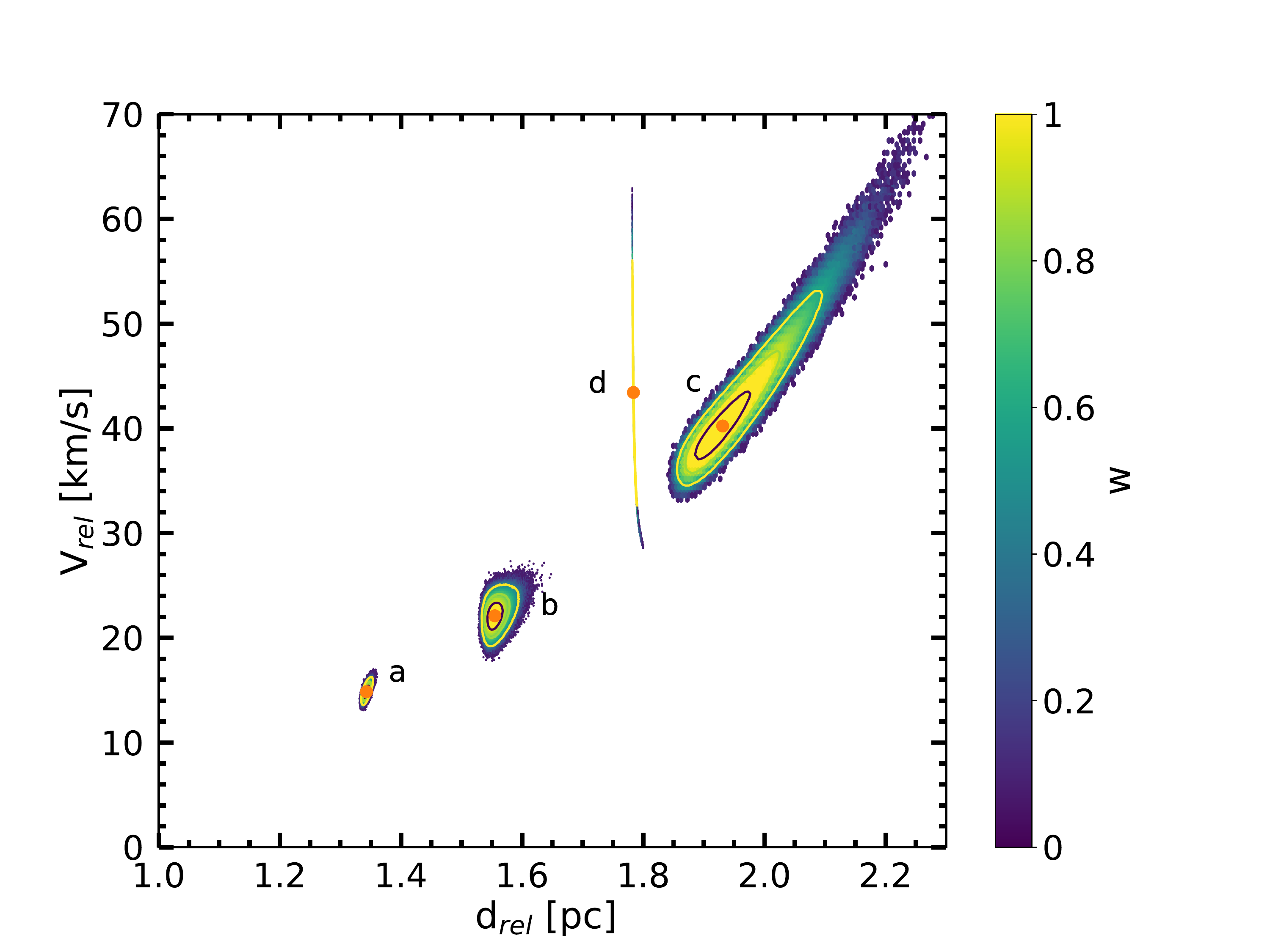}
\caption{Probability density-distribution in relative distance
  ($d_{\rm rel}$) and relative velocity ($v_{\rm rel}$) between \A and
  the four closest stars $a$, $b$, $c$ and $d$ (see
  Table\,\ref{Tab:ClosestCandidates}). The colour bar to the right gives
  a weigthing ($w$) with respect to the $A$ and $v$ from
  Eq.\,\ref{Eq:Volume} and Eq.\,\ref{Eq:Crosssection} with respect to
  the same values of \A.
   \label{Fig:Gaia_A_dv}
}
\end{center}
\end{figure}

\begin{figure}
\begin{center}
\includegraphics[width=1.0\columnwidth]{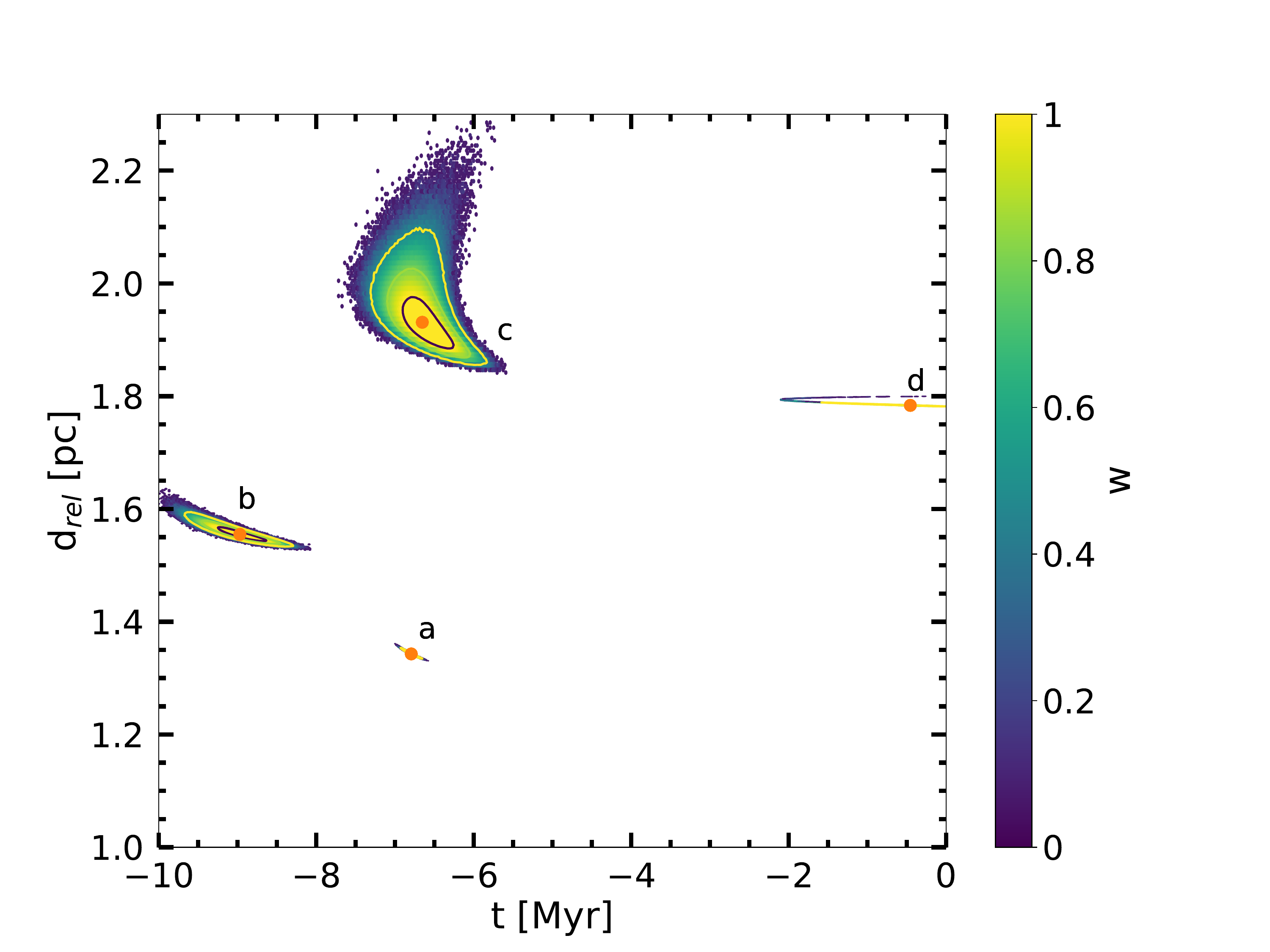}
\caption{Probability density-distribution in time of closest approach
  ($t$) and relative distance ($d_{\rm rel}$) between \A and the four
  closest stars (see Table\,\ref{Tab:ClosestCandidates}).
  \label{Fig:Gaia_A_td}
}
\end{center}
\end{figure}

We now generate one million objects within the error ellipsoid (in
astrometric and radial velocity) of each of the selected stars. This
results, for each star, in a probability density-distribution in time,
relative distance and relative velocity of the closest approach
between that particular star and \A. In Fig.\,\ref{Fig:Gaia_A_dv} and
Fig.\,\ref{Fig:Gaia_A_td} we show the various probability
distributions for time of closest approach, relative distance, and
relative velocity for the 4 stars within 2\,pc that have a closest
approach.  We identify these stars in
Table\,\ref{Tab:ClosestCandidates}.  About 6.8\,Myr ago, \A passed the
star HIP\,17288 within a distance of $\sim 1.3$\,pc and with a
relative velocity of $\sim 15$\,km/s which is consistent with the
result of \cite{2018ApJ...852L..27F} and \cite{2018A&A...610L..11D}.
We find several other relatively close encounters which, despite their
close proximity with \A, we have not included here because of the
large uncertainty in the radial velocity.  Based on this data, we do
not expect that \A was launched from any of the nearby stars in the
Gaia catalogue.  We speculate that \A originates from well beyond the
solar neighbourhood.

\begin{table*}
  \begin{center}
  \caption{Closest star encounters with \A within 2 pc.  Columns 1 to
    3 give the identity of the star, the following columns give the
    time of the encounter, the closest relative distance with \A,
    relative velocity, and the variation intervals of the samples in
    time, distance and velocity. Note that time is negative,
    indicating that the encounter happened in the past.
 \medskip
 \label{Tab:ClosestCandidates}}
 \begin{tabular}{lcllllllr}
 \hline
 \hline
  ID & Gaia identity &HIP&$t$ & $t_{\rm sample}$  & $d_{\rm rel}$&  $d_{\rm sample}$ & $v_{\rm rel}$ & $v_{\rm sample}$\\
         &&&$[Myr]$&  & $[pc]$ & & $[km/s]$ \\
 \hline
$^{a}$ &5108377030337405952& 17288 & -6.793 & [-7.011, -6.574]& 1.342 &[1.330, 1.363] & 14.854& [12.917, 17.184]\\
$^{b}$ &4863923915804133376& & -8.970 &[-10.135, -8.037] &1.554 &[1.526, 1.644]&  22.106 &[17.281, 27.255]\\
$^{c}$ &5140501942602437632& &-6.653& [-7.755, -5.481]   &1.931& [1.840, 2.447] &40.228 &[32.951, 85.899] \\ 
$^{d}$ &1362592668307182592& 86916& -0.455& [-2.083, 2.223] & 1.783 &[1.781, 1.799] & 43.430 &[28.745, 62.525]\\
\hline
\hline
 \end{tabular}
 \end{center}
\end{table*}

\subsection{An origin from beyond the solar neighbourhood}

\begin{figure}
\begin{center}
\includegraphics[width=1.0\columnwidth]{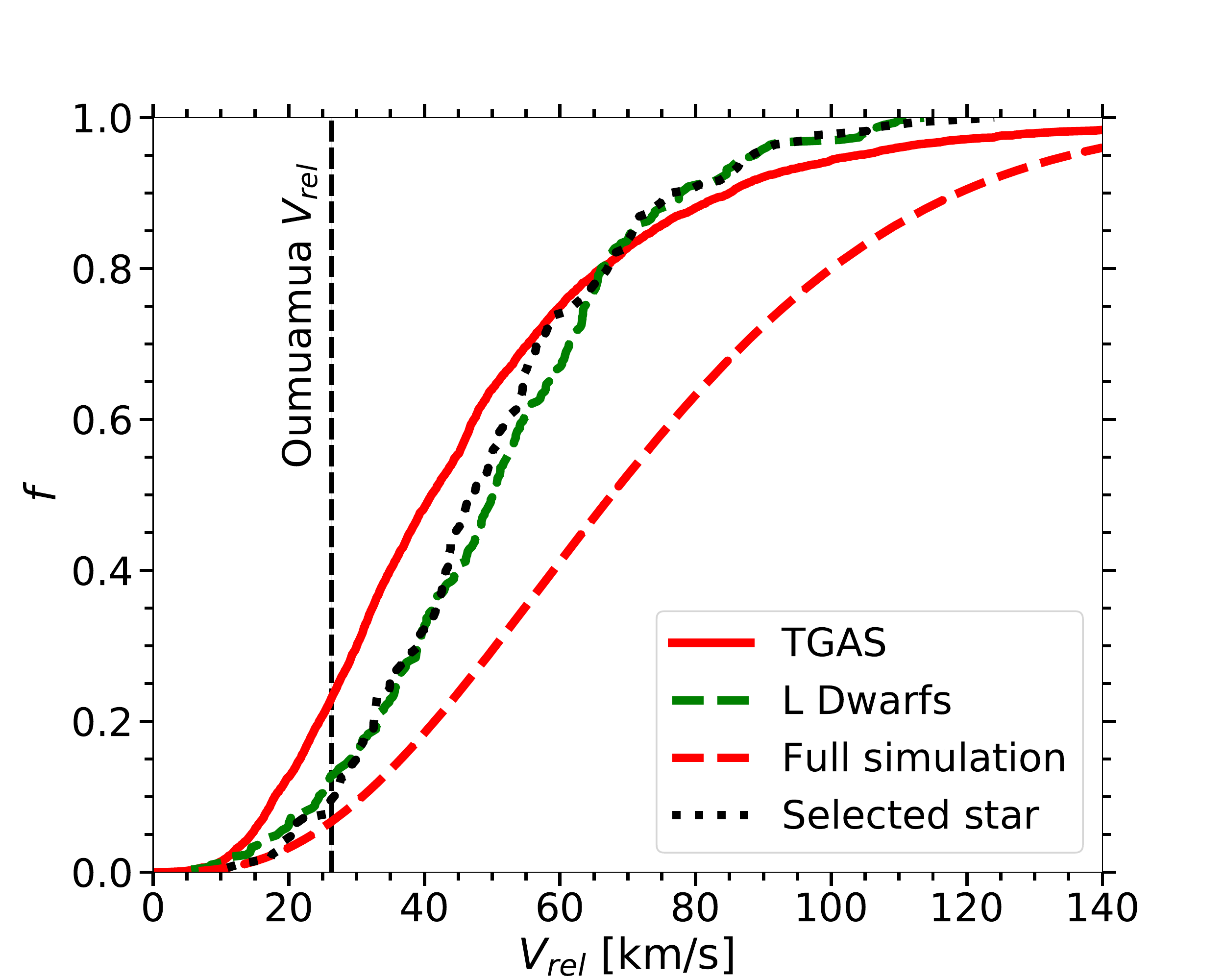}
\caption{Cumulative relative velocity distributions of single objects
  in the solar neighbourhood.  \A is presented as the vertical line
  near 26\,km/s.  The solid red curve gives the velocity distribution
  of stars within 30\,pc of the Sun from the Gaia-TGAS catalogues.  The
  green dashed curve gives the distribution of 100 L-dwarfs within
  $\sim 20$\,pc of the Sun \citep{2010AJ....139.1808S}.  The red
  dashed curve gives the mean relative velocity-distribution of all
  stars within 7.5-8.5\,kpc from the Galactic centre, whereas the
  black dotted curve gives the distribution between one Sun-like star
  \citep[at a distance of $\sim 7.2$\,kpc from the centre with a
    velocity of $\sim 223.1$\,km/s and at an angle of $\sim 20^\circ$
    from the tip of the bar,][]{2009A&A...501..941H} and its
  neighbours.  The latter distribution is statistically consistent
  with that of the brown dwarfs with KS-probability of $\sim 0.3$.
  \label{Fig:velocity_distribution}
}
\end{center}
\end{figure}

If \A is part of the Galactic background distribution of \soli{}, its
velocity is expected to be consistent with the distribution of
low-mass objects in the Galaxy.  To analyse quantitatively this
hypothesis, we compare the velocity of \A with the distribution of the
stars in Gaia-TGAS. This distribution is presented in
Fig.\,\ref{Fig:velocity_distribution}.

Ideally, we would like to compare \A to a population of objects with
similar characteristics.  We decide to take the population of
L-dwarfs, which in mass are sufficiently small that they could be
considered mass-less in the Galactic potential. In addition, there is
a reasonable consistent census of the population of L-dwarfs
within 20\,pc of the solar neighbourhood \citep{2012ApJ...753..156K}.
For these dwarfs, the velocity has been measured as a function of age
\citep{2015ApJS..220...18B}, and this distribution follows:
\begin{equation}
  v \simeq {t \over [{\rm Gyr}]} \times 6~~{\rm km/s} + 20~~{\rm
    km/s}.
\end{equation}
Inverting this relation leads to an age estimate for \A of $\sim
1$\,Gyr, which is consistent with the sample of youngest
($1.1$--$1.7$\,Gyr) L-dwarfs \citep{2015ApJS..220...18B}.  Considering
the velocity distribution of the entire population of brown dwarfs
results in an average age of about 5\,Gyr (see
Fig.\,\ref{Fig:velocity_distribution}), which is consistent with the
estimate of $5.2\pm 0.2$\,Gyr by \cite{2015ApJS..220...18B}, whereas
for the Gaia-TGAS data, we find $3\pm0.5$\,Gyr. The difference between
the brown-dwarf age-estimates and those in the Gaia-TGAS catalogue is
possibly due to a selection effect caused by the magnitude-limited
sample in the latter \citep[see also][]{2018A&A...609A...8B}.

To further validate the use of the brown-dwarf velocity-distribution
to constrain the age of \A, we compare it to the velocity distribution
of objects in recent simulations of the Milky Way Galaxy by
\cite{2017arXiv171200058F}.  This calculation, performed with the {\tt
  Bonsai} \citep{2012JCoPh.231.2825B, 2014hpcn.conf...54B}
gravitational tree-code using a shared time-step of $\sim 0.6$ Myr, a
gravitational softening length of $10$\,pc and with opening angle
$\theta=0.4$ are carried out with $8 \times 10^9$ particles in a
stellar disc and include a live halo, bulge, and bar.  In this
snapshot at an age of 10\,Gyr, we selected objects within 30\,pc
around a hypothetical star that matches the current relative location
of the Sun in the Milky Way.  At this distance, the structure of the
disc is no longer visible and the inclination distribution is
flat. The resulting relative velocity distribution of the full sample
is presented as the red-dashed curve and a sample star that matches
closest to the L-dwarfs is represented by the black-dotted curve in
Fig.\,\ref{Fig:velocity_distribution}.

The age distribution of all the objects in the simulations is
consistent with an age of $\sim 8.3$\,Gyr, but for the star
representing the Sun it is comparable to the age-velocity distribution
of 100 L-dwarfs in the age-range of 1.1--1.7\,Gyr of
\cite{2018MNRAS.475.1093Y}.  Instead of comparing the relative
velocity-distributions directly, they should be weighted by the
encounter rate and corrected for gravitational focusing (see
Eq.\,\ref{Eq:Crosssection}). For the relevant velocity range ($v_{\rm
  rel} \aplt 100$\,km/s) however, this correction is small ($< 25$\%).

\section{Discussion and conclusions}

Based on the existence of \A, we derive a local density of $3.5 \times
10^{13}$--$2.1 \times 10^{15}$\,pc$^{-3}$
(0.0040--0.24\,au$^{-3}$).  This is a high density, but in line
with other estimates \citep{2018ApJ...855L..10D}. It is consistent
with the amount of debris ejected during the star and planet formation
process, but inconsistent with expulsion of exo Oort-cloud or
asteroidal objects (see \S\,\ref{Sect:Aisrare}).

If each star contributes to the formation of \soli{} then the entire
Galaxy may be swarming with such objects, with {\cal O}($10^{23}$)
\soli{} in the Milky Way.  Once liberated from their parent stars, it
is quite possible that such an object grazes any other star, much in
the same way \A passed close to the Sun. The velocity of \A is low
compared to the stars in the Gaia-TGAS catalogue, and low compared to
the mean velocity distribution of L-dwarfs. Young L-dwarfs have a
considerably lower velocity than their older siblings, and the
velocity of \A is consistent with the youngest population of brown
dwarfs (1.1--1.7\,Gyr).  Based on its velocity, we argue that \A has a
similar age \citep[see also][]{2018ApJ...852L..27F}.

Earlier estimates of the interstellar asteroid density were carried
out to explain the daily X-ray flares on the super-massive black-hole
in the Galactic centre Sgr A$\star$ \citep{2015MNRAS.446..710H}.  They
argue that the daily X-ray flares in the direction of Sgr A$\star$ can
be explained with a local density of $\sim 10^{14}$ asteroids
pc$^{-3}$, which is consistent with the density of \soli{} derived
here.

We conclude that \A~is part of the left-over debris of the star and
planet formation process in the Galaxy. We expect that the Galaxy is
rich in such objects, with a density of $\sim 10^{14}$ or $10^{15}$
objects per cubic parsec.  We estimate the probability that a \solus{}
passes the Sun within 1\,au, taking the gravitational focusing
corrected cross-section into account, at an event rate of about 2--12
per year.

{\bf Acknowledgements} We thank Fabo Feng for discussions on a
possible origin of \A in the Oort cloud.  We thank Ben Pole, Yohai
Meiron and Maura Portegies Zwart for the translation of the term
``lonely rock'' into Latin.
 We thank the referees for their thorough
 checking of facts, the corrections to our interpretation of the MNRAS
 style documents, and lessons in the subtle differences between British
 and American English.
This work was supported by the Netherlands
Research School for Astronomy (NOVA) and NWO (grant \#621.016.701
[LGM-II]). This work was supported by a grant from the Swiss National
Supercomputing Centre (CSCS) under project ID s716.  This work used
the public git version of the {\tt AMUSE} software environment which
can be found at \url{https://github.com/amusecode}.  This work has
made use of data from the European Space Agency (ESA) mission Gaia
(\url{https://www.cosmos. esa.int/gaia}), processed by the Gaia Data
Processing and Analysis Consortium (DPAC,
\url{https://www.cosmos.esa. int/web/gaia/dpac/consortium}). Funding
for the DPAC has been provided by national institutions, in particular
the institutions participating in the Gaia Multilateral Agreement.

\input{Rama.bbl}
\label{lastpage}

\begin{thebibliography}{}
\makeatletter
\relax
\def\mn@urlcharsother{\let\do\@makeother \do\$\do\&\do\#\do\^\do\_\do\%\do\~}
\def\mn@doi{\begingroup\mn@urlcharsother \@ifnextchar [ {\mn@doi@}
  {\mn@doi@[]}}
\def\mn@doi@[#1]#2{\def\@tempa{#1}\ifx\@tempa\@empty \href
  {http://dx.doi.org/#2} {doi:#2}\else \href {http://dx.doi.org/#2} {#1}\fi
  \endgroup}
\def\mn@eprint#1#2{\mn@eprint@#1:#2::\@nil}
\def\mn@eprint@arXiv#1{\href {http://arxiv.org/abs/#1} {{\tt arXiv:#1}}}
\def\mn@eprint@dblp#1{\href {http://dblp.uni-trier.de/rec/bibtex/#1.xml}
  {dblp:#1}}
\def\mn@eprint@#1:#2:#3:#4\@nil{\def\@tempa {#1}\def\@tempb {#2}\def\@tempc
  {#3}\ifx \@tempc \@empty \let \@tempc \@tempb \let \@tempb \@tempa \fi \ifx
  \@tempb \@empty \def\@tempb {arXiv}\fi \@ifundefined
  {mn@eprint@\@tempb}{\@tempb:\@tempc}{\expandafter \expandafter \csname
  mn@eprint@\@tempb\endcsname \expandafter{\@tempc}}}

\bibitem[\protect\citeauthoryear{{Antoja} et~al.,}{{Antoja}
  et~al.}{2014}]{2014A&A...563A..60A}
{Antoja} T.,  et~al., 2014, \mn@doi [\aap] {10.1051/0004-6361/201322623}, \href
  {http://adsabs.harvard.edu/abs/2014A%26A...563A..60A} {563, A60}

\bibitem[\protect\citeauthoryear{{Bacci} et~al.,}{{Bacci}
  et~al.}{2017}]{2017MPEC....U..181B}
{Bacci} P.,  et~al., 2017, Minor Planet Electronic Circulars, 2017

\bibitem[\protect\citeauthoryear{{Bailer-Jones}}{{Bailer-Jones}}{2018}]{2018A&A...609A...8B}
{Bailer-Jones} C.~A.~L.,  2018, \mn@doi [\aap] {10.1051/0004-6361/201731453},
  \href {http://adsabs.harvard.edu/abs/2018A%26A...609A...8B} {609, A8}

\bibitem[\protect\citeauthoryear{{Bannister} et~al.,}{{Bannister}
  et~al.}{2017}]{2017ApJ...851L..38B}
{Bannister} M.~T.,  et~al., 2017, \mn@doi [\apjl] {10.3847/2041-8213/aaa07c},
  \href {http://adsabs.harvard.edu/abs/2017ApJ...851L..38B} {851, L38}

\bibitem[\protect\citeauthoryear{{B{\'e}dorf}, {Gaburov}, {Fujii}, {Nitadori},
  {Ishiyama}  \& {Portegies Zwart}}{{B{\'e}dorf}
  et~al.}{2014}]{2014hpcn.conf...54B}
{B{\'e}dorf} J.,  {Gaburov} E.,  {Fujii} M.~S.,  {Nitadori} K.,  {Ishiyama} T.,
    {Portegies Zwart} S.,  2014, in Proceedings of the International Conference
  for High Performance Computing, Networking, Storage and Analysis, p. 54-65.
  pp 54--65 (\mn@eprint {arXiv} {1412.0659}), \mn@doi{10.1109/SC.2014.10OTHER:
  http://dl.acm.org/citation.cfm?id=2683600}

\bibitem[\protect\citeauthoryear{{B{\'e}dorf}, {Gaburov}  \& {Portegies
  Zwart}}{{B{\'e}dorf} et~al.}{2012}]{2012JCoPh.231.2825B}
{B{\'e}dorf} J.,  {Gaburov} E.,   {Portegies Zwart} S.,  2012, \mn@doi [Journal
  of Computational Physics] {10.1016/j.jcp.2011.12.024}, \href
  {http://adsabs.harvard.edu/abs/2012JCoPh.231.2825B} {231, 2825}
  
\bibitem[\protect\citeauthoryear{{Boersma}}{{Boersma}}{1961}]{1961BAN....15..291B}
{Boersma} J.,  1961, \bain, \href
  {http://adsabs.harvard.edu/abs/1961BAN....15..291B} {15, 291}

\bibitem[\protect\citeauthoryear{{Bolin} et~al.,}{{Bolin}
  et~al.}{2018}]{2018ApJ...852L...2B}
{Bolin} B.~T.,  et~al., 2018, \mn@doi [\apjl] {10.3847/2041-8213/aaa0c9}, \href
  {http://adsabs.harvard.edu/abs/2018ApJ...852L...2B} {852, L2}

\bibitem[\protect\citeauthoryear{{Brasser} \& {Morbidelli}}{{Brasser} \&
  {Morbidelli}}{2013}]{2013Icar..225...40B}
{Brasser} R.,  {Morbidelli} A.,  2013, \mn@doi [\icarus]
  {10.1016/j.icarus.2013.03.012}, \href
  {http://adsabs.harvard.edu/abs/2013Icar..225...40B} {225, 40}

\bibitem[\protect\citeauthoryear{{Brown}}{{Brown}}{2005}]{2005IAUC.8636....1B}
{Brown} M.~E.,  2005, \iaucirc, \href
  {http://adsabs.harvard.edu/abs/2005IAUC.8636....1B} {8636}

\bibitem[\protect\citeauthoryear{{Brown} \& {Butler}}{{Brown} \&
  {Butler}}{2017}]{2017AJ....154...19B}
{Brown} M.~E.,  {Butler} B.~J.,  2017, \mn@doi [\aj]
  {10.3847/1538-3881/aa6346}, \href
  {http://adsabs.harvard.edu/abs/2017AJ....154...19B} {154, 19}

\bibitem[\protect\citeauthoryear{{Burgasser} et~al.,}{{Burgasser}
  et~al.}{2015}]{2015ApJS..220...18B}
{Burgasser} A.~J.,  et~al., 2015, \mn@doi [\apjs] {10.1088/0067-0049/220/1/18},
  \href {http://adsabs.harvard.edu/abs/2015ApJS..220...18B} {220, 18}

\bibitem[\protect\citeauthoryear{{Chabrier}}{{Chabrier}}{2003}]{2003PASP..115..763C}
{Chabrier} G.,  2003, \mn@doi [\pasp] {10.1086/376392}, \href
  {http://adsabs.harvard.edu/abs/2003PASP..115..763C} {115, 763}

\bibitem[\protect\citeauthoryear{{{\'C}uk}}{{{\'C}uk}}{2018}]{2018ApJ...852L..15C}
{{\'C}uk} M.,  2018, \mn@doi [\apjl] {10.3847/2041-8213/aaa3db}, \href
  {http://adsabs.harvard.edu/abs/2018ApJ...852L..15C} {852, L15}

\bibitem[\protect\citeauthoryear{{de la Fuente Marcos} \& {de la Fuente
  Marcos}}{{de la Fuente Marcos} \& {de la Fuente
  Marcos}}{2017}]{2017RNAAS...1....5D}
{de la Fuente Marcos} C.,  {de la Fuente Marcos} R.,  2017, \mn@doi [Research
  Notes of the American Astronomical Society] {10.3847/2515-5172/aa96b4}, \href
  {http://adsabs.harvard.edu/abs/2017RNAAS...1....5D} {1, 5}

\bibitem[\protect\citeauthoryear{{de la Fuente Marcos}, {de la Fuente Marcos}
  \& {Aarseth}}{{de la Fuente Marcos} et~al.}{2017}]{2017Ap&SS.362..198D}
{de la Fuente Marcos} C.,  {de la Fuente Marcos} R.,   {Aarseth} S.~J.,  2017,
  \mn@doi [\apss] {10.1007/s10509-017-3181-1}, \href
  {http://adsabs.harvard.edu/abs/2017Ap\%26SS.362..198D} {362, 198}

\bibitem[\protect\citeauthoryear{{de la Fuente Marcos}, {de la Fuente Marcos}
  \& {Aarseth}}{{de la Fuente Marcos} et~al.}{2018}]{2018MNRAS.476L...1D}
{de la Fuente Marcos} C.,  {de la Fuente Marcos} R.,   {Aarseth} S.~J.,  2018,
  \mn@doi [\mnras] {10.1093/mnrasl/sly019}, \href
  {http://adsabs.harvard.edu/abs/2018MNRAS.476L...1D} {476, L1}

\bibitem[\protect\citeauthoryear{{Do}, {Tucker}  \& {Tonry}}{{Do}
  et~al.}{2018}]{2018ApJ...855L..10D}
{Do} A.,  {Tucker} M.~A.,   {Tonry} J.,  2018, \mn@doi [\apjl]
  {10.3847/2041-8213/aaae67}, \href
  {http://adsabs.harvard.edu/abs/2018ApJ...855L..10D} {855, L10}

\bibitem[\protect\citeauthoryear{{Domokos}, {Sipos}, {Szab{\'o}}  \&
  {V{\'a}rkonyi}}{{Domokos} et~al.}{2017}]{2017RNAAS...1...50D}
{Domokos} G.,  {Sipos} A.~{\'A}.,  {Szab{\'o}} G.~M.,   {V{\'a}rkonyi} P.~L.,
  2017, \mn@doi [Research Notes of the American Astronomical Society]
  {10.3847/2515-5172/aaa12f}, \href
  {http://adsabs.harvard.edu/abs/2017RNAAS...1...50D} {1, 50}

\bibitem[\protect\citeauthoryear{{Dybczy{\'n}ski} \&
  {Kr{\'o}likowska}}{{Dybczy{\'n}ski} \&
  {Kr{\'o}likowska}}{2018}]{2018A&A...610L..11D}
{Dybczy{\'n}ski} P.~A.,  {Kr{\'o}likowska} M.,  2018, \mn@doi [\aap]
  {10.1051/0004-6361/201732309}, \href
  {http://adsabs.harvard.edu/abs/2018A%26A...610L..11D} {610, L11}

\bibitem[\protect\citeauthoryear{{Engelhardt}, {Jedicke}, {Vere{\v s}},
  {Fitzsimmons}, {Denneau}, {Beshore}  \& {Meinke}}{{Engelhardt}
  et~al.}{2017}]{2017AJ....153..133E}
{Engelhardt} T.,  {Jedicke} R.,  {Vere{\v s}} P.,  {Fitzsimmons} A.,  {Denneau}
  L.,  {Beshore} E.,   {Meinke} B.,  2017, \mn@doi [\aj]
  {10.3847/1538-3881/aa5c8a}, \href
  {http://adsabs.harvard.edu/abs/2017AJ....153..133E} {153, 133}

\bibitem[\protect\citeauthoryear{{Feng} \& {Jones}}{{Feng} \&
  {Jones}}{2018}]{2018ApJ...852L..27F}
{Feng} F.,  {Jones} H.~R.~A.,  2018, \mn@doi [\apjl]
  {10.3847/2041-8213/aaa404}, \href
  {http://adsabs.harvard.edu/abs/2018ApJ...852L..27F} {852, L27}

\bibitem[\protect\citeauthoryear{{Ferrin} \& {Zuluaga}}{{Ferrin} \&
  {Zuluaga}}{2017}]{2017arXiv171107535F}
{Ferrin} I.,  {Zuluaga} J.,  2017, preprint, \href
  {http://adsabs.harvard.edu/abs/2017arXiv171107535F} {} (\mn@eprint {arXiv}
  {1711.07535})

\bibitem[\protect\citeauthoryear{{Fitzsimmons} et~al.,}{{Fitzsimmons}
  et~al.}{2018}]{2018NatAs...2..133F}
{Fitzsimmons} A.,  et~al., 2018, \mn@doi [Nature Astronomy]
  {10.1038/s41550-017-0361-4}, \href
  {http://adsabs.harvard.edu/abs/2018NatAs...2..133F} {2, 133}

\bibitem[\protect\citeauthoryear{{Fouchard}, {Rickman}, {Froeschl{\'e}}  \&
  {Valsecchi}}{{Fouchard} et~al.}{2014}]{2014Icar..231...99F}
{Fouchard} M.,  {Rickman} H.,  {Froeschl{\'e}} C.,   {Valsecchi} G.~B.,  2014,
  \mn@doi [\icarus] {10.1016/j.icarus.2013.11.034}, \href
  {http://adsabs.harvard.edu/abs/2014Icar..231...99F} {231, 99}

\bibitem[\protect\citeauthoryear{{Francis}}{{Francis}}{2005}]{2005ApJ...635.1348F}
{Francis} P.~J.,  2005, \mn@doi [\apj] {10.1086/497684}, \href
  {http://adsabs.harvard.edu/abs/2005ApJ...635.1348F} {635, 1348}

\bibitem[\protect\citeauthoryear{{Fraser} et~al.,}{{Fraser}
  et~al.}{2017}]{2017NatAs...1E..88F}
{Fraser} W.~C.,  et~al., 2017, \mn@doi [Nature Astronomy]
  {10.1038/s41550-017-0088}, \href
  {http://adsabs.harvard.edu/abs/2017NatAs...1E..88F} {1, 0088}

\bibitem[\protect\citeauthoryear{{Fujii}, {B{\'e}dorf}, {Baba}  \& {Portegies
  Zwart}}{{Fujii} et~al.}{2017}]{2017arXiv171200058F}
{Fujii} M.~S.,  {B{\'e}dorf} J.,  {Baba} J.,   {Portegies Zwart} S.,  2017,
  preprint, \href {http://adsabs.harvard.edu/abs/2017arXiv171200058F} {}
  (\mn@eprint {arXiv} {1712.00058})

\bibitem[\protect\citeauthoryear{{Gaia Collaboration} et~al.,}{{Gaia
  Collaboration} et~al.}{2016}]{2016A&A...595A...2G}
{Gaia Collaboration} et~al., 2016, \mn@doi [\aap]
  {10.1051/0004-6361/201629512}, \href
  {http://adsabs.harvard.edu/abs/2016A%26A...595A...2G} {595, A2}

\bibitem[\protect\citeauthoryear{{Gaidos}}{{Gaidos}}{2017}]{2017arXiv171206721G}
{Gaidos} E.,  2017, preprint, \href
  {http://adsabs.harvard.edu/abs/2017arXiv171206721G} {} (\mn@eprint {arXiv}
  {1712.06721})

\bibitem[\protect\citeauthoryear{{Gaidos}, {Williams}  \& {Kraus}}{{Gaidos}
  et~al.}{2017}]{2017RNAAS...1...13G}
{Gaidos} E.,  {Williams} J.,   {Kraus} A.,  2017, \mn@doi [Research Notes of
  the American Astronomical Society] {10.3847/2515-5172/aa9851}, \href
  {http://adsabs.harvard.edu/abs/2017RNAAS...1...13G} {1, 13}

\bibitem[\protect\citeauthoryear{{Gomes}, {Levison}, {Tsiganis}  \&
  {Morbidelli}}{{Gomes} et~al.}{2005}]{2005Natur.435..466G}
{Gomes} R.,  {Levison} H.~F.,  {Tsiganis} K.,   {Morbidelli} A.,  2005, \mn@doi
  [\nat] {10.1038/nature03676}, \href
  {http://adsabs.harvard.edu/abs/2005Natur.435..466G} {435, 466}

\bibitem[\protect\citeauthoryear{{Gontcharov}}{{Gontcharov}}{2006}]{2006AstL...32..759G}
{Gontcharov} G.~A.,  2006, \mn@doi [Astronomy Letters]
  {10.1134/S1063773706110065}, \href
  {http://adsabs.harvard.edu/abs/2006AstL...32..759G} {32, 759}

\bibitem[\protect\citeauthoryear{{Hamers} \& {Portegies Zwart}}{{Hamers} \&
  {Portegies Zwart}}{2015}]{2015MNRAS.446..710H}
{Hamers} A.~S.,  {Portegies Zwart} S.~F.,  2015, \mn@doi [\mnras]
  {10.1093/mnras/stu2103}, \href
  {http://adsabs.harvard.edu/abs/2015MNRAS.446..710H} {446, 710}

\bibitem[\protect\citeauthoryear{{Hanse}, {J{\'{\i}}lkov{\'a}}, {Portegies
  Zwart}  \& {Pelupessy}}{{Hanse} et~al.}{2018}]{2018MNRAS.473.5432H}
{Hanse} J.,  {J{\'{\i}}lkov{\'a}} L.,  {Portegies Zwart} S.~F.,   {Pelupessy}
  F.~I.,  2018, \mn@doi [\mnras] {10.1093/mnras/stx2721}, \href
  {http://adsabs.harvard.edu/abs/2018MNRAS.473.5432H} {473, 5432}

\bibitem[\protect\citeauthoryear{{Hansen} \& {Zuckerman}}{{Hansen} \&
  {Zuckerman}}{2017}]{2017RNAAS...1...55H}
{Hansen} B.,  {Zuckerman} B.,  2017, \mn@doi [Research Notes of the American
  Astronomical Society] {10.3847/2515-5172/aaa3ee}, \href
  {http://adsabs.harvard.edu/abs/2017RNAAS...1...55H} {1, 55}

\bibitem[\protect\citeauthoryear{{Hills}}{{Hills}}{1988}]{1988Natur.331..687H}
{Hills} J.~G.,  1988, \mn@doi [\nat] {10.1038/331687a0}, \href
  {http://adsabs.harvard.edu/abs/1988Natur.331..687H} {331, 687}

\bibitem[\protect\citeauthoryear{{Hoang}, {Loeb}, {Lazarian}  \& {Cho}}{{Hoang}
  et~al.}{2018}]{2018arXiv180201335H}
{Hoang} T.,  {Loeb} A.,  {Lazarian} A.,   {Cho} J.,  2018, preprint, \href
  {http://adsabs.harvard.edu/abs/2018arXiv180201335H} {} (\mn@eprint {arXiv}
  {1802.01335})

\bibitem[\protect\citeauthoryear{{Holmberg}, {Nordstr{\"o}m}  \&
  {Andersen}}{{Holmberg} et~al.}{2009}]{2009A&A...501..941H}
{Holmberg} J.,  {Nordstr{\"o}m} B.,   {Andersen} J.,  2009, \mn@doi [\aap]
  {10.1051/0004-6361/200811191}, \href
  {http://adsabs.harvard.edu/abs/2009A%26A...501..941H} {501, 941}

\bibitem[\protect\citeauthoryear{{Jackson}, {Tamayo}, {Hammond}, {Ali-Dib}  \&
  {Rein}}{{Jackson} et~al.}{2017}]{2017arXiv171204435J}
{Jackson} A.~P.,  {Tamayo} D.,  {Hammond} N.,  {Ali-Dib} M.,   {Rein} H.,
  2017, preprint, \href {http://adsabs.harvard.edu/abs/2017arXiv171204435J} {}
  (\mn@eprint {arXiv} {1712.04435})

\bibitem[\protect\citeauthoryear{{J{\"a}nes}, {Pelupessy}  \& {Portegies
  Zwart}}{{J{\"a}nes} et~al.}{2014}]{2014A&A...570A..20J}
{J{\"a}nes} J.,  {Pelupessy} I.,   {Portegies Zwart} S.,  2014, \mn@doi [\aap]
  {10.1051/0004-6361/201423831}, \href
  {http://adsabs.harvard.edu/abs/2014A\%26A...570A..20J} {570, A20}

\bibitem[\protect\citeauthoryear{{Jewitt}}{{Jewitt}}{2018}]{2018AJ....155...56J}
{Jewitt} D.,  2018, \mn@doi [\aj] {10.3847/1538-3881/aaa1a4}, \href
  {http://adsabs.harvard.edu/abs/2018AJ....155...56J} {155, 56}

\bibitem[\protect\citeauthoryear{{Jewitt}, {Luu}, {Rajagopal}, {Kotulla},
  {Ridgway}, {Liu}  \& {Augusteijn}}{{Jewitt}
  et~al.}{2017}]{2017ApJ...850L..36J}
{Jewitt} D.,  {Luu} J.,  {Rajagopal} J.,  {Kotulla} R.,  {Ridgway} S.,  {Liu}
  W.,   {Augusteijn} T.,  2017, \mn@doi [\apjl] {10.3847/2041-8213/aa9b2f},
  \href {http://adsabs.harvard.edu/abs/2017ApJ...850L..36J} {850, L36}

\bibitem[\protect\citeauthoryear{{J{\'{\i}}lkov{\'a}}, {Hamers}, {Hammer}  \&
  {Portegies Zwart}}{{J{\'{\i}}lkov{\'a}} et~al.}{2016}]{2016MNRAS.457.4218J}
{J{\'{\i}}lkov{\'a}} L.,  {Hamers} A.~S.,  {Hammer} M.,   {Portegies Zwart} S.,
   2016, \mn@doi [\mnras] {10.1093/mnras/stw264}, \href
  {http://adsabs.harvard.edu/abs/2016MNRAS.457.4218J} {457, 4218}

\bibitem[\protect\citeauthoryear{{Kaib} \& {Quinn}}{{Kaib} \&
  {Quinn}}{2009}]{2009Sci...325.1234K}
{Kaib} N.~A.,  {Quinn} T.,  2009, \mn@doi [Science] {10.1126/science.1172676},
  \href {http://adsabs.harvard.edu/abs/2009Sci...325.1234K} {325, 1234}

\bibitem[\protect\citeauthoryear{{Katz}}{{Katz}}{2018}]{2018arXiv180202273K}
{Katz} J.~I.,  2018, preprint, \href
  {http://adsabs.harvard.edu/abs/2018arXiv180202273K} {} (\mn@eprint {arXiv}
  {1802.02273})

\bibitem[\protect\citeauthoryear{{Kipper}, {Tempel}  \& {Tenjes}}{{Kipper}
  et~al.}{2018}]{2018MNRAS.473.2188K}
{Kipper} R.,  {Tempel} E.,   {Tenjes} P.,  2018, \mn@doi [\mnras]
  {10.1093/mnras/stx2441}, \href
  {http://adsabs.harvard.edu/abs/2018MNRAS.473.2188K} {473, 2188}

\bibitem[\protect\citeauthoryear{{Kirkpatrick} et~al.,}{{Kirkpatrick}
  et~al.}{2012}]{2012ApJ...753..156K}
{Kirkpatrick} J.~D.,  et~al., 2012, \mn@doi [\apj]
  {10.1088/0004-637X/753/2/156}, \href
  {http://adsabs.harvard.edu/abs/2012ApJ...753..156K} {753, 156}

\bibitem[\protect\citeauthoryear{{Knight}, {Protopapa}, {Kelley}, {Farnham},
  {Bauer}, {Bodewits}, {Feaga}  \& {Sunshine}}{{Knight}
  et~al.}{2017}]{2017ApJ...851L..31K}
{Knight} M.~M.,  {Protopapa} S.,  {Kelley} M.~S.~P.,  {Farnham} T.~L.,  {Bauer}
  J.~M.,  {Bodewits} D.,  {Feaga} L.~M.,   {Sunshine} J.~M.,  2017, \mn@doi
  [\apjl] {10.3847/2041-8213/aa9d81}, \href
  {http://adsabs.harvard.edu/abs/2017ApJ...851L..31K} {851, L31}

\bibitem[\protect\citeauthoryear{{Kuiper}}{{Kuiper}}{1951}]{1951PNAS...37....1K}
{Kuiper} G.~P.,  1951, \mn@doi [Proceedings of the National Academy of Science]
  {10.1073/pnas.37.1.1}, \href
  {http://adsabs.harvard.edu/abs/1951PNAS...37....1K} {37, 1}

\bibitem[\protect\citeauthoryear{{Kunder} et~al.,}{{Kunder}
  et~al.}{2017}]{2017AJ....153...75K}
{Kunder} A.,  et~al., 2017, \mn@doi [\aj] {10.3847/1538-3881/153/2/75}, \href
  {http://adsabs.harvard.edu/abs/2017AJ....153...75K} {153, 75}

\bibitem[\protect\citeauthoryear{{Mamajek}}{{Mamajek}}{2017}]{2017RNAAS...1...21M}
{Mamajek} E.,  2017, \mn@doi [Research Notes of the American Astronomical
  Society] {10.3847/2515-5172/aa9bdc}, \href
  {http://adsabs.harvard.edu/abs/2017RNAAS...1...21M} {1, 21}

\bibitem[\protect\citeauthoryear{{Mart{\'{\i}}nez-Barbosa}, {Brown}  \&
  {Portegies Zwart}}{{Mart{\'{\i}}nez-Barbosa}
  et~al.}{2015}]{2015MNRAS.446..823M}
{Mart{\'{\i}}nez-Barbosa} C.~A.,  {Brown} A.~G.~A.,   {Portegies Zwart} S.,
  2015, \mn@doi [\mnras] {10.1093/mnras/stu2094}, \href
  {http://adsabs.harvard.edu/abs/2015MNRAS.446..823M} {446, 823}

\bibitem[\protect\citeauthoryear{{Masiero}}{{Masiero}}{2017}]{2017arXiv171009977M}
{Masiero} J.,  2017, preprint, \href
  {http://adsabs.harvard.edu/abs/2017arXiv171009977M} {} (\mn@eprint {arXiv}
  {1710.09977})

\bibitem[\protect\citeauthoryear{{McGlynn} \& {Chapman}}{{McGlynn} \&
  {Chapman}}{1989}]{1989ApJ...346L.105M}
{McGlynn} T.~A.,  {Chapman} R.~D.,  1989, \mn@doi [\apjl] {10.1086/185590},
  \href {http://adsabs.harvard.edu/abs/1989ApJ...346L.105M} {346, L105}

\bibitem[\protect\citeauthoryear{{Meech}}{{Meech}}{2018}]{2018NatAs...2..112M}
{Meech} K.~J.,  2018, \mn@doi [Nature Astronomy] {10.1038/s41550-018-0382-7},
  \href {http://adsabs.harvard.edu/abs/2018NatAs...2..112M} {2, 112}

\bibitem[\protect\citeauthoryear{{Meech} et~al.,}{{Meech}
  et~al.}{2017a}]{2017Natur.552..378M}
{Meech} K.~J.,  et~al., 2017a, \mn@doi [\nat] {10.1038/nature25020}, \href
  {http://adsabs.harvard.edu/abs/2017Natur.552..378M} {552, 378}

\bibitem[\protect\citeauthoryear{{Meech} et~al.,}{{Meech}
  et~al.}{2017b}]{2017MPEC....U..183M}
{Meech} K.~J.,  et~al., 2017b, Minor Planet Electronic Circulars, 2017

\bibitem[\protect\citeauthoryear{{Meech} et~al.,}{{Meech}
  et~al.}{2017c}]{2017MPEC....W...52M}
{Meech} K.~J.,  et~al., 2017c, Minor Planet Electronic Circulars, 2017

\bibitem[\protect\citeauthoryear{{Moro-Mart{\'{\i}}n}, {Turner}  \&
  {Loeb}}{{Moro-Mart{\'{\i}}n} et~al.}{2009}]{2009ApJ...704..733M}
{Moro-Mart{\'{\i}}n} A.,  {Turner} E.~L.,   {Loeb} A.,  2009, \mn@doi [\apj]
  {10.1088/0004-637X/704/1/733}, \href
  {http://adsabs.harvard.edu/abs/2009ApJ...704..733M} {704, 733}

\bibitem[\protect\citeauthoryear{{Napiwotzki}}{{Napiwotzki}}{2009}]{2009JPhCS.172a2004N}
{Napiwotzki} R.,  2009, in Journal of Physics Conference Series. p. 012004
  (\mn@eprint {arXiv} {0903.2159}), \mn@doi{10.1088/1742-6596/172/1/012004}

\bibitem[\protect\citeauthoryear{{Noll}, {Grundy}, {Chiang}, {Margot}  \&
  {Kern}}{{Noll} et~al.}{2008}]{2008ssbn.book..345N}
{Noll} K.~S.,  {Grundy} W.~M.,  {Chiang} E.~I.,  {Margot} J.-L.,   {Kern}
  S.~D.,  2008, {Binaries in the Kuiper Belt}.
pp 345--363

\bibitem[\protect\citeauthoryear{{Oort}}{{Oort}}{1950}]{1950BAN....11...91O}
{Oort} J.~H.,  1950, \bain, \href
  {http://adsabs.harvard.edu/abs/1950BAN....11...91O} {11, 91}

\bibitem[\protect\citeauthoryear{{Pelupessy}, {van Elteren}, {de Vries},
  et~al.}{{Pelupessy}
  et~al.}{2013}]{2013AA...557A..84P}
{Pelupessy} F.~I.,  {van Elteren} A.,  {de Vries} N.,  {McMillan} S.~L.~W.,
  {Drost} N.,   {Portegies Zwart} S.~F.,  2013, \mn@doi [\aap]
  {10.1051/0004-6361/201321252}, \href
  {http://adsabs.harvard.edu/abs/2013A%26A...557A..84P} {557, A84}

\bibitem[\protect\citeauthoryear{{Portegies Zwart} \& {McMillan}}{{Portegies
  Zwart} \& {McMillan}}{2017}]{AMUSE}
{Portegies Zwart} S.,  {McMillan} S.,  2017, {Astrophysical Recipes: the Art of
  AMUSE}.
AAS IOP Astronomy

\bibitem[\protect\citeauthoryear{{Portegies Zwart}, {McMillan}, {van Elteren},
  et~al.}{{Portegies Zwart}
  et~al.}{2013}]{2013CoPhC.184..456P}
{Portegies Zwart} S.~F.,  {McMillan} S.~L.~W.,  {van Elteren} A.,  {Pelupessy}
  F.~I.,   {de Vries} N.,  2013, \mn@doi [Computer Physics Communications]
  {10.1016/j.cpc.2012.09.024}, \href
  {http://adsabs.harvard.edu/abs/2013CoPhC.184..456P} {184, 456}

\bibitem[\protect\citeauthoryear{{Rafikov}}{{Rafikov}}{2018}]{2018arXiv180102658R}
{Rafikov} R.~R.,  2018, preprint, \href
  {http://adsabs.harvard.edu/abs/2018arXiv180102658R} {} (\mn@eprint {arXiv}
  {1801.02658})

\bibitem[\protect\citeauthoryear{{Raymond}, {Veras}, {Armitage}, {Barclay}  \&
  {Quintana}}{{Raymond} et~al.}{2018a}]{2018MNRAS.tmp..493R}
{Raymond} S.~N.,  {Veras} D.,  {Armitage} P.~J.,  {Barclay} T.,   {Quintana}
  E.,  2018a, \mn@doi [\mnras] {10.1093/mnras/sty468}, \href
  {http://adsabs.harvard.edu/abs/2018MNRAS.tmp..493R} {}

\bibitem[\protect\citeauthoryear{{Raymond}, {Armitage}  \& {Veras}}{{Raymond}
  et~al.}{2018b}]{2018arXiv180302840R}
{Raymond} S.~N.,  {Armitage} P.~J.,   {Veras} D.,  2018b, preprint, \href
  {http://adsabs.harvard.edu/abs/2018arXiv180302840R} {} (\mn@eprint {arXiv}
  {1803.02840})

\bibitem[\protect\citeauthoryear{{Rickman}, {Fouchard}, {Froeschl{\'e}}  \&
  {Valsecchi}}{{Rickman} et~al.}{2008}]{2008CeMDA.102..111R}
{Rickman} H.,  {Fouchard} M.,  {Froeschl{\'e}} C.,   {Valsecchi} G.~B.,  2008,
  \mn@doi [Celestial Mechanics and Dynamical Astronomy]
  {10.1007/s10569-008-9140-y}, \href
  {http://adsabs.harvard.edu/abs/2008CeMDA.102..111R} {102, 111}

\bibitem[\protect\citeauthoryear{{Schlichting}, {Fuentes}  \&
  {Trilling}}{{Schlichting} et~al.}{2013}]{2013AJ....146...36S}
{Schlichting} H.~E.,  {Fuentes} C.~I.,   {Trilling} D.~E.,  2013, \mn@doi [\aj]
  {10.1088/0004-6256/146/2/36}, \href
  {http://adsabs.harvard.edu/abs/2013AJ....146...36S} {146, 36}

\bibitem[\protect\citeauthoryear{{Schmidt}, {West}, {Hawley}  \&
  {Pineda}}{{Schmidt} et~al.}{2010}]{2010AJ....139.1808S}
{Schmidt} S.~J.,  {West} A.~A.,  {Hawley} S.~L.,   {Pineda} J.~S.,  2010,
  \mn@doi [\aj] {10.1088/0004-6256/139/5/1808}, \href
  {http://adsabs.harvard.edu/abs/2010AJ....139.1808S} {139, 1808}

\bibitem[\protect\citeauthoryear{{Shepard} et~al.,}{{Shepard}
  et~al.}{2017}]{2017Icar..281..388S}
{Shepard} M.~K.,  et~al., 2017, \mn@doi [\icarus]
  {10.1016/j.icarus.2016.08.011}, \href
  {http://adsabs.harvard.edu/abs/2017Icar..281..388S} {281, 388}

\bibitem[\protect\citeauthoryear{{Sheppard} \& {Jewitt}}{{Sheppard} \&
  {Jewitt}}{2004}]{2004AJ....127.3023S}
{Sheppard} S.~S.,  {Jewitt} D.,  2004, \mn@doi [\aj] {10.1086/383558}, \href
  {http://adsabs.harvard.edu/abs/2004AJ....127.3023S} {127, 3023}

\bibitem[\protect\citeauthoryear{{Stern}}{{Stern}}{1990}]{1990PASP..102..793S}
{Stern} S.~A.,  1990, \mn@doi [\pasp] {10.1086/132704}, \href
  {http://adsabs.harvard.edu/abs/1990PASP..102..793S} {102, 793}

\bibitem[\protect\citeauthoryear{{Torres}, {Brown}  \& {Portegies
  Zwart}}{{Torres} et~al.}{2017}]{2017IAUS..330T}
{Torres} S.,  {Brown} A.~G.~A.,   {Portegies Zwart} S.-F.,  2017, in
  Proceedings IAU Symposium.

\bibitem[\protect\citeauthoryear{{Trilling} et~al.,}{{Trilling}
  et~al.}{2017}]{2017ApJ...850L..38T}
{Trilling} D.~E.,  et~al., 2017, \mn@doi [\apjl] {10.3847/2041-8213/aa9989},
  \href {http://adsabs.harvard.edu/abs/2017ApJ...850L..38T} {850, L38}

\bibitem[\protect\citeauthoryear{{Veras}, {Wyatt}, {Mustill}, {Bonsor}  \&
  {Eldridge}}{{Veras} et~al.}{2011}]{2011MNRAS.417.2104V}
{Veras} D.,  {Wyatt} M.~C.,  {Mustill} A.~J.,  {Bonsor} A.,   {Eldridge} J.~J.,
   2011, \mn@doi [\mnras] {10.1111/j.1365-2966.2011.19393.x}, \href
  {http://adsabs.harvard.edu/abs/2011MNRAS.417.2104V} {417, 2104}

\bibitem[\protect\citeauthoryear{{Widmark} \& {Monari}}{{Widmark} \&
  {Monari}}{2017}]{2017arXiv171107504W}
{Widmark} A.,  {Monari} G.,  2017, preprint, \href
  {http://adsabs.harvard.edu/abs/2017arXiv171107504W} {} (\mn@eprint {arXiv}
  {1711.07504})

\bibitem[\protect\citeauthoryear{{Wright}}{{Wright}}{2017}]{2017RNAAS...1...38W}
{Wright} J.~T.,  2017, \mn@doi [Research Notes of the American Astronomical
  Society] {10.3847/2515-5172/aa9f23}, \href
  {http://adsabs.harvard.edu/abs/2017RNAAS...1...38W} {1, 38}

\bibitem[\protect\citeauthoryear{{Ye}, {Zhang}, {Kelley}  \& {Brown}}{{Ye}
  et~al.}{2017}]{2017ApJ...851L...5Y}
{Ye} Q.-Z.,  {Zhang} Q.,  {Kelley} M.~S.~P.,   {Brown} P.~G.,  2017, \mn@doi
  [\apjl] {10.3847/2041-8213/aa9a34}, \href
  {http://adsabs.harvard.edu/abs/2017ApJ...851L...5Y} {851, L5}

\bibitem[\protect\citeauthoryear{{Yu} \& {Liu}}{{Yu} \&
  {Liu}}{2018}]{2018MNRAS.475.1093Y}
{Yu} J.,  {Liu} C.,  2018, \mn@doi [\mnras] {10.1093/mnras/stx3204}, \href
  {http://adsabs.harvard.edu/abs/2018MNRAS.475.1093Y} {475, 1093}

\bibitem[\protect\citeauthoryear{{Zhang}}{{Zhang}}{2018}]{2018ApJ...852L..13Z}
{Zhang} Q.,  2018, \mn@doi [\apjl] {10.3847/2041-8213/aaa2f7}, \href
  {http://adsabs.harvard.edu/abs/2018ApJ...852L..13Z} {852, L13}

\makeatother
\end{thebibliography}
\end{document}